\documentclass[fleqn,twoside,twocolumn,nofootinbib,showkeys]{revtex4} 
\usepackage[sec,nocpr]{ujp} 

\begin{document}
\title[Features of Charge Transport in Mo/$n$-Si
Structures]
{FEATURES OF CHARGE TRANSPORT IN Mo/\boldmath$n$-Si
STRUCTURES WITH A SCHOTTKY BARRIER}%
\author{O.Ya. Olikh}
\affiliation{Taras Shevchenko National University of Kyiv, Faculty of Physics}
\address{64, Volodymyrs'ka Str., Kyiv 01601, Ukraine}
\email{olikh@univ.kiev.ua}
\udk{537.312.6/.7:621.314.63}  \razd{\secvii}

\autorcol{O.Ya.\hspace*{0.7mm}Olikh}

\setcounter{page}{126}%

\begin{abstract}
Forward and reverse current-voltage characteristics of Mo/$n$-Si
Schottky barrier structures have been studied experimentally in the
temperature range 130\,$\div $\,330~K. The Schottky barrier height
is found to increase and the ideality factor to decrease, as the
temperature grows. The obtained results are analyzed in the
framework of a non-uniform contact model. The average value and the
standard deviation of a Schottky barrier height are determined to be
0.872 and 0.099~V, respectively, at $T=$~130\,$\div $\,220~K and
0.656 and 0.036~V, respectively, at $T=$ =~230\,$\div $\,330~K.
Thermionic emission over the non-uniform barrier and tunneling are
shown to be the dominant processes of charge transfer at a reverse
bias voltage.
\end{abstract}


\maketitle

\section{Introduction}

Structures with a Schottky contact are widely used, while
manufacturing high-speed logic, integrated, and opto-electronic
elements.
Therefore, the interest of scientists to similar structures is quite
obvious.
One of the basic approaches to the description of a current through the
the metal--semiconductor (MS) contact is the theory of thermionic emission
(TE).
In an ideal case, the TE current $I$ through the structure is described
by
the expression \cite{Colinge,Rhoderick,Striha}
\begin{equation}
I=I_{\rm S}\{\exp [qV/(kT)]-1\},  \label{1m}
\end{equation}
where $V$ is the bias voltage applied to the structure,
\begin{equation}
I_{\rm S}=SA^{\ast }T^{2}\exp [-q\Phi _{b}/(kT)]  \label{2m}
\end{equation}%
is the saturation current at the reverse bias, $S$ is the contact
area, $A^{\ast }$ is the effective Richardson constant, and $\Phi
_{b}$ is the Schottky barrier height (SBH). The latter is defined as
the difference between metal's work function and the semiconductor
electron affinity \cite{Colinge}. However, expression (\ref{1m}) is
too simplified for the case of real MS structures, for which such
phenomena as the action of image forces, the presence of an
intermediate dielectric layer and electron states at the interface,
a non-uniformity of contact, and a drop of the applied voltage not
only across the depletion region in the semiconductor have to be
taken into account. As a consequence, the following equation is
often used to describe the TE current through the Schottky
contact~\cite{Rhoderick}:
\begin{equation}
I=I_{\rm S}\exp \!\left[ \frac{q(V-IR_{\rm S})}{nkT}\right] \!
\left\{\! 1-\exp \left[ -\frac{q(V-IR_{\rm S})}{kT}\right]
\!\right\}\!, \label{3m}
\end{equation}%
where $n$ is the ideality factor, $R_{\rm S}$ is the series
resistance, and $I_{\rm S}$ is also described by expression
(\ref{2m}), but the quantity $\Phi _{b}$, as well as $n$, becomes
dependent on the contact state and the temperature $T$. Besides TE,
probable mechanisms of charge transfer in MS structures are
generation-recombination processes in the contact region, various
current leakage processes, tunneling, thermally induced field
emission (local energy levels can play a considerable role in the
last two cases), and so forth
\cite{Rhoderick,Arslan,Donoval,Huang,Evstropov,Lee,Sathaiyaa,Hsu}.
As a result, the total current is often considered as a sum of
several terms. Each of them is associated with a specific charge
transfer mechanism, which can dominate in that or another
temperature or field range \cite{Arslan,Donoval,Huang}. In
particular, in many cases--such as generation-recombination
currents, thermally induced field emission, trap-assisted tunneling,
and so on--those terms look similar to those in Eq.~(\ref{3m}).
However, while describing the saturation current, other expressions,
different from Eq.~(\ref{2m}), have to be applied.

Note that, owing to a wide variety of affecting factors, the problem of
predicting a specific charge transfer mechanism in structures with a
Schottky barrier under definite conditions, including temperature ones,
is a
complicated task, which has no general solution. On the other hand, the
technology development assumes that the scope of requirements should be
extended to include the conditions, under which semiconductor devices would
operate. Therefore, the aim of this work was to elucidate the mechanisms
of
charge transfer at forward and reverse biases in Mo/$n$-Si structures,
which
were fabricated following the standard industrial technology, at
temperatures
below their nominal operating range. Analogous structures are used at
manufacturing the rectifying diodes, in particular, of 2D219 type. As a
result, the measurement of current-voltage characteristics (CVCs) was
selected as the main method of researches. The data obtained are
analyzed
in the framework of the inhomogeneous barrier model
\cite{Werner,Sullivan,Tung}.
Lately, this model has been used more and more widely for the
interpretation
of experimental data obtained for various structures with a Schottky barrier different by their
composition
\cite{Sarpatwari,Biber,Tascioglu,Yildirim,Mamor,Iucolano,Iucolano2}.

\begin{figure}
\includegraphics[width=\column]{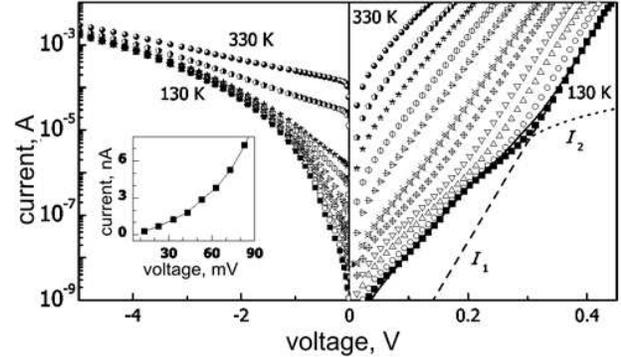}
\vskip-3mm\caption{Forward (right panel) and reverse (left panel)
CVC branches for Mo/$n $-Si Schottky diodes measured in the
temperature range 130--330~K with an increment of
20$~\mathrm{K}$. The curves in the right panel illustrate the
approximation of the forward CVC branch at $T=130~\mathrm{K}$ by formula
(\ref{4m}): current $I_{1}$ (dashed curve), current $I_{2}$ (dotted
curve), and their sum (solid curve); the corresponding approximation
parameters are $n_{1}=1.67$, $n_{2}=2.53$, $I_{\rm S1}=5.0\times
10^{-13}$~A, $I_{\rm S2}=3.8\times 10^{-10}$~A, and $R_{\rm
S}=4.1\times 10^{3}$~$\Omega $. The initial section of the forward CVC
branch at $T=130$~K is shown in the inset  }\label{Fig1}
\end{figure}

\section{Specimen Fabrication. Measurement and~Calculation Techniques}

In our researches, we used Schottky diodes with the following structure.
An
epitaxial $n$-Si:P layer 0.2~$\mu \mathrm{m}$ in thickness was deposited
on
a $n^{+}$-Si:Sb substrate (KES~0.01, a thickness of 250~$\mu
\mathrm{m}$). A
Schottky contact 2~mm in diameter was created on the epitaxial layer
surface
by depositing a molybdenum layer. The other, ohmic, contact was created
on
the opposite substrate side. The structures were fabricated at the
Tomilino
Electronic Factory (Russia).

The current-voltage characteristics of the structures concerned were
measured in the interval of dc current variation of
$10^{-9}$--$10^{-2}$~A at the forward and reverse biases with a
voltage increment of 0.01~V and in the temperature interval
130--330$~\mathrm{K}$. The specimen temperature was monitored with the
use of a copper-constantan thermocouple.

Some examples of forward and reverse CVC branches registered at various
temperatures are depicted in Fig.~\ref{Fig1}. One can see that, at
temperatures higher than 250$~\mathrm{K}$, the forward CVC branches are
almost linear on the semilogarithmic scale, within the interval of
current
variation being of about three orders of magnitude. At the same time, at
$T<210$~K, the total current can be divided into two components. In
particular, for the CVC associated with the current that dominates at
low
biases, the influence of the series resistance is substantial, which is
evidenced by a deviation of the exhibited curves from linearity in the
interval $7\times 10^{-8}~\mathrm{A}<I<5\times 10^{-7}~\mathrm{A}$. In
this
connection and taking Eq.~(\ref{3m}) into account, the following
expression
was used to describe the forward CVC branches:
\[
I=I_{1}+I_{2}=I_{\rm S1}\exp {\left(\! \frac{qV}{n_{1}kT}\!\right)\!
}\left[ 1-\exp {\left( \!-\frac{qV}{kT}\!\right) }\right] +
\]
\begin{equation}
+I_{\rm S2}\exp\! \left[ \frac{q(V\!-\!IR_{\rm
S})}{n_{2}kT}\right]\! \left\{\! 1\!-\!\exp \left[
-\frac{q(V\!-\!IR_{\rm S})}{kT}\right] \!\right\}\!. \label{4m}
\end{equation}%
Here, the first term prevails at $I>10^{-5}$~A, and the second one at
$I<5\times 10^{-7}$~A. Note that another known technique used to make
allowance for the existence of CVC peculiarities at low biases consists
in
the insertion of a shunting resistance rather than the term $I_{2}$.
However, in our opinion, such an approach is not justified in this case,
because the forward CVC branch is not linear even at the lowest biases
(see
the inset in Fig.~1).

\begin{figure}
\includegraphics[width=\column]{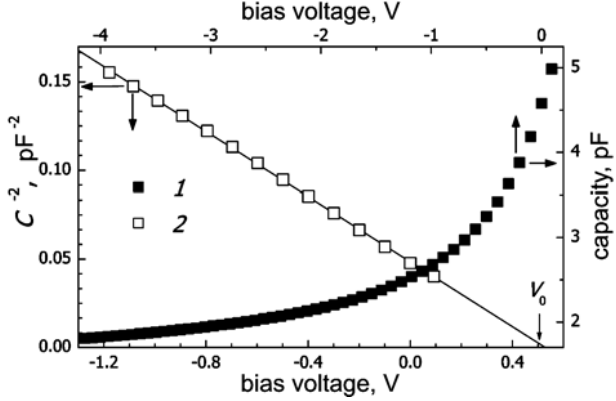}
\vskip-3mm\caption{Dependence of the capacity $C$ (curve \textit{1})
and the quantity $C^{-2}$ (curve \textit{2}) on the applied voltage
for Mo/$n$-Si Schottky diodes at $T=295~\mathrm{K}$. Points
correspond to experimental data, and the curve to their linear
approximation }\label{Fig2}\vskip2mm
\end{figure}

In order to determine the fitting parameters, the following
procedure was used. Two sections were selected in the forward CVC
branch, in which $10^{-5}~\mathrm{A}<$ $<I<10^{-2}~\mathrm{A}$ and
$10^{-9}~\mathrm{A}<I<10^{-7}~\mathrm{A}$, respectively. The data
for the former were used to plot the dependence of the quantity $\ln
{{I}/{\left[ 1-\exp \left( -qV/kT\right) \right] }}$ on $V$. The
obtained curve was approximated by a straight line, the slope and
the free term of which were related to the magnitudes of $n_{1} $
and $I_{\rm S1}$, respectively. From the data for the latter section
and using the Cheung \cite{Cheung} and Gromov \cite{Gromov} methods,
the magnitude of $R_{\rm S}$ was determined. The application of two
techniques was aimed at enhancing the reliability of data obtained.
The corresponding values turned out to be equal to each other to
within 10\%. After the value for $R_{\rm S}$ had been determined,
the quantity $V$ in the latter section was replaced by the effective
voltage $V^{\ast }=V-IR_{\rm S}$. Then the $I_{\rm S2}$- and
$n_{2}$-values were determined following the procedure described
above.

Figure~\ref{Fig1} illustrates an example of the approximation of the
experimental
forward CVC branch registered at a certain temperature. The
approximation
was carried out with the use of formula (\ref{4m}), and the parameters
were obtained following the described routine. A good coincidence
between
the calculated curve and the experimental points is evident.

Basing on expression (\ref{2m}) and obtained $I_{\rm S1}$- and
$I_{\rm S2}$-values, we also determined the corresponding SBHs at
the zero bias, $\Phi _{b1}$ and $\Phi _{b2}$, respectively. In the
calculations, we assumed that, for $n$-Si, $A^{\ast }=$
$=112$~A/$\mathrm{cm}^{\mathrm{2}}/\mathrm{K}^{2}$ \cite{Schroder}
and $S=3.14\times 10^{-6}$~m$^{2}$.

For monitoring the doping level, we measured the capacity-voltage
(volt-farad) characteristics (VFCs) of the studied structures at
room temperature, $T=$ $=295~\mathrm{K}$; see Fig.~\ref{Fig2}. The
results obtained show that the concentration of charge carriers in
the epitaxial layer is $N_{D}=1.3\times 10^{23}$~m$^{-3}$. In
addition, with the help of the expression \cite{Rhoderick,Schroder}
\begin{equation}
\Phi _{b,{\rm CV}}=V_{n}+V_{0}+kT/q,  \label{8am}
\end{equation}%
we determined the SBH $\Phi _{b,{\rm CV}}=(0.689\pm 0.002)$~V. In
Eq.~(\ref{8am}), the quantity $qV_{n}=kT\ln {(N_{C}/N_{D})}$ equals
the energy difference between the conduction band bottom and the
Fermi level position, $N_{C}$ is the effective density of states
near the conduction band bottom, and $V_{0}$ is the abscissa of
intersection point between the voltage axis and a straight line
approximating the dependence of $C^{-2}$, where $C$ is the capacity
of a Schottky diode, on the reverse bias voltage. Note that the SBH
determined in such a way should exceed the corresponding value
obtained with the use of CVCs \cite{Rhoderick}.

\section{Results and Their Discussion}

\begin{figure}
\includegraphics[width=7.5cm]{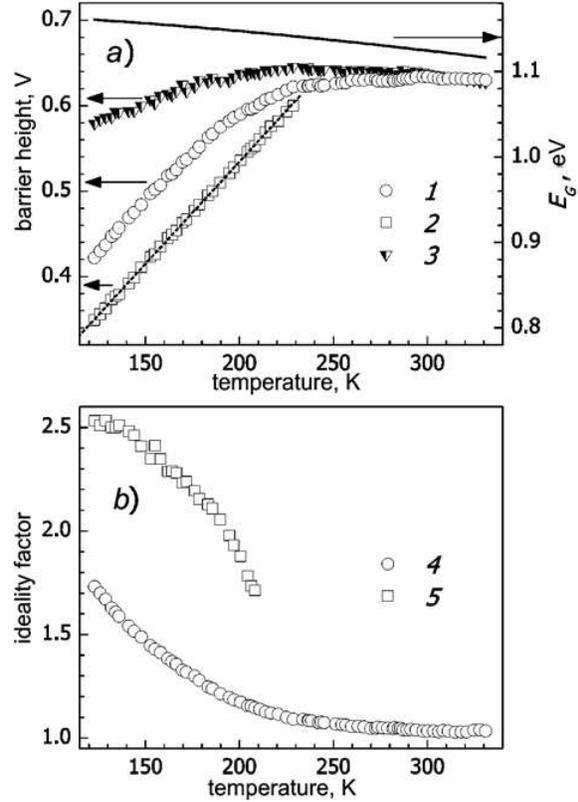}
\vskip-3mm\caption{Temperature dependences of the barrier height
(\textit{a}) and the ideality factor (\textit{b}) for Mo/$n$-Si
Schottky diodes: $\Phi _{b1}$ (\textit{1}), $\Phi _{b2}$
(\textit{2}), $\Phi _{\rm bef}$ (\textit{3}), $n_{1}$ (\textit{4}),
and $n_{2}$ (\textit{5}). The dotted line depicts the linear
approximation of curve \textit{2}. The solid curve in panel $a$
corresponds to the temperature dependence of the energy gap width in
Si } \label{Fig3}\vskip4mm
\end{figure}

First, let us consider the features in the current that prevails at
high temperatures and high biases; it is $I_{1}$. The obtained
temperature dependences of parameters are shown in Fig.~\ref{Fig3}.
One can see that, the value of $\Phi _{b1}$ increases, as the
temperature grows. It was experimentally demonstrated
\cite{Zhu,Aboelfotoh} that the opposite tendency has to be observed
at the temperature elevation in real structures with a uniform
Schottky barrier, provided that the TE dominates, with the
temperature coefficients for the reduction of the SBH and the energy
gap width $E_{G}$ being very close to each other in this case. On
the other hand, it is known that the SBH determined with the help of
CVCs can differ from the real one. In particular, the authors of
work \cite{Bozhkov} assert the necessity of carrying out
measurements at a constant current across the contact and propose to
use the expression
\begin{equation}
\Phi _{\rm bef}=n_{I_{C}}\Phi _{b}-(n_{I_{C}}-1) ({kT}/{q}) \ln
\left( {SA^{\ast }T^{2}}/{I_{C}}\right),   \label{9m}
\end{equation}%
where $n_{I_{C}}$ is the ideality factor at a definite constant
current $I_{C}$, for the evaluation of the effective barrier height
$\Phi _{\rm bef}$. In the cited work \cite{Bozhkov}, it was shown
that, in the case of TE through a homogeneous contact, the magnitude
of $\Phi _{\rm bef}$ almost coincides with the real barrier height
and the both quantities have the same temperature dependence.

We calculated $\Phi_{\rm bef}$ according to formula (\ref{9m}) at
$I_{C}=10^{-3}$~A (Fig.~\ref{Fig3}, curve {\it 3}). For the sake of
comparison, the same figure exhibits the temperature dependence of
$E_{G}$. When calculating the latter, we took
$E_{G}(T)=E_{G}(0)-\gamma T^{2}/(T+\beta)$, where
$E_{G}(0)=1.17$~eV, $\beta=636$~K, and
$\gamma=4.73\times10^{-4}$~eV$/$K$^{2}$~\cite{Markvart}. One can see
that, although the magnitude of $\Phi_{\rm bef}$ varies within a
much narrower interval, its temperature dependence also differs from
the behavior of $E_{G}$, especially at low temperatures.

\begin{figure}
\includegraphics[width=\column]{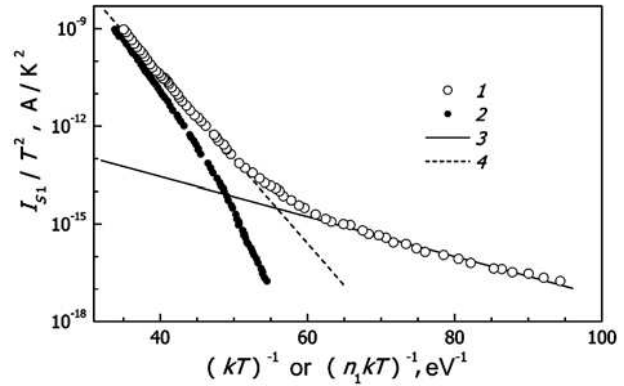}
\vskip-3mm\caption{Ordinary (\textit{1}) and transformed
(\textit{2}) Richardson dependences for $ I _{1}$. The straight
lines are linear approximations of the data in curve \textit{1} in
the intervals $T=(130\div220)~\mathrm{K}$ (\textit{3}) and
$T=(230\div330)~\mathrm{K}$ (\textit{4}) }\label{Fig4}
\end{figure}

On the other hand, there exists a procedure to calculate the
parameter $A^{\ast}$ \cite{Rhoderick,Schroder} consisting in the
plotting of the Richardson dependence, i.e. the dependences of the
quantity $\ln(I_{\rm S}/T^{2}) $ on $(kT)^{-1}$ (see
Fig.~\ref{Fig4}, curve~{\it 1}). According to Eq.~(\ref{2m}), it has
to be described by the expression
\begin{equation}
\ln\left( {I_{\rm S}}/{T^{2}}\right)
=\ln(SA^{\ast})-{q\Phi_{b}}/{(kT)}. \label{10m}
\end{equation}
However, it is evident that the linear dependence is really
observed, but only in two intervals rather than in the whole
temperature range. The calculations by formula (\ref{10m}) gave rise
to $\Phi_{bR,{\rm I}}=(0.141\pm0.004) $~V and $A_{R,{\rm
I}}^{\ast}=(3.7\pm0.8)\times10^{-10}$~A/cm$^{2}/$K$^{2}$ in the
interval 130--220$~\mathrm{K}$ and to $\Phi_{bR,{\rm
II}}=(0.599\pm0.003)$~V and $A_{R,{\rm
II}}^{\ast}=(30\pm10)$~A/cm$^{2}/$K$^{2}$ in the interval
230--330$~\mathrm{K}$. It is evident that the values of $A_{R,{\rm
II}}^{\ast}$ and, especially, $A_{R,{\rm I}}^{\ast}$ differ from the
literature data.

As we know from the literature \cite{Schmitsdorf}, in the case of a
substantial deviation from ideality, it is expedient to use the
transformed Richardson dependence for the determination of
$A^{\ast}$, in which the quantity $(nkT)^{-1}$ rather than
$(kT)^{-1}$ is reckoned along the abscissa axis. However, in our
case, the transformed Richardson dependence (Fig.~\ref{Fig4}, curve
{\it 2}) is also \mbox{non-linear.}

\begin{figure}
\includegraphics[width=\column]{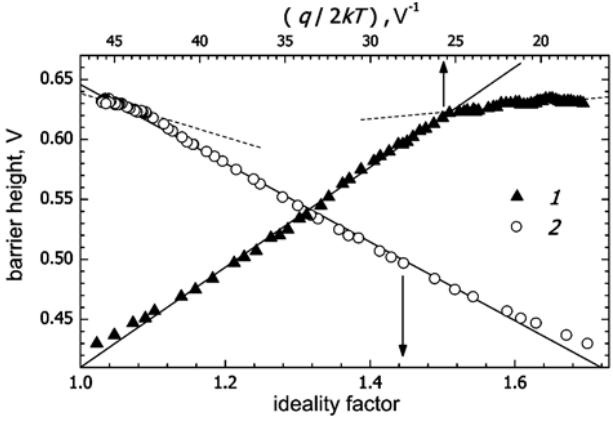}
\vskip-3mm\caption{Dependences of $\Phi_{b1}$-value on the
reciprocal of the doubled temperature (\textit{1}) and $n_{1}$
(\textit{2}). The straight lines are linear approximations in the
intervals $T=(130\div220)~\mathrm{K}$ (solid lines) and
$T=(230\div330)~\mathrm{K}$ (dotted lines)  }\label{Fig5}\vskip3mm
\end{figure}

\begin{figure}
\includegraphics[width=\column]{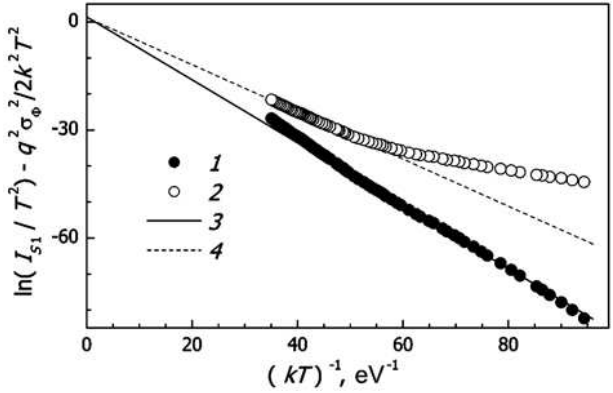}
\vskip-3mm\caption{Modified Richardson dependences (\ref{16m}) for
$I_{\rm S1}$. $\sigma _{0}=0.099$ (\textit{1}) and 0.036~V
(\textit{2}). Straight lines \textit{3} and \textit{4} are the
linear approximations of curve~\textit{1} in the interval
$T=(130\div220)~\mathrm{K}$ and curve~\textit{2} in the interval
$T=(230\div330)~\mathrm{K}$, respectively  }\label{Fig6}
\end{figure}

While summarizing the above consideration, it is necessary to recognize that the
obtained results cannot be explained in the framework of the theory for
TE
through a uniform contact. On the other hand, the model of inhomogeneous
Schottky barrier has often been used recently to explain the CVCs of
real
metal--semiconductor structures~\cite{Werner,Sullivan,Tung}. In
particular,
according to the model proposed in work \cite{Werner}, if the SBH is
described by a Gaussian distribution, the theory of TE brings about the
relation~\cite{Werner,Tascioglu,Yildirim,Mamor}
\begin{equation}
\Phi _{b}=\Phi _{b}^{0}-{q\sigma _{\Phi }^{2}}/{(2kT)},  \label{13m}
\end{equation}%
where $\Phi _{b}^{0}$ is the average SBH, and $\sigma _{\Phi }$ is
the standard deviation of a barrier height, which characterizes the
contact uniformity. The corresponding dependence for our case is
depicted in Fig.~\ref{Fig5} (curve {\it 1}). One can see that the
linear dependence really takes place, however, in two separate
temperature intervals, $T=$ $=130\div 220$ and $230\div
330~\mathrm{K}$. Linear approximations by formula (\ref{13m}) gave
the following fitting parameters: $\Phi _{bT,{\rm I}}^{0}=(0.872\pm
0.003)$~V and $\sigma _{\Phi,{\rm I}}=(0.099\pm 0.001)$~V for the
former interval, and $\Phi _{bT,{\rm II}}^{0}=(0.656\pm 0.003)$~V
and $\sigma _{\Phi,{\rm II}}=(0.036\pm 0.004)$~V for the latter one.

In the framework of another model for the non-uniform Schottky
contact~\cite{Sullivan,Tung}, the SBH is assumed identical over the
whole MS interface,
excluding the sections with small areas (patches), where the SBH is lower.
The
patches may differ from one another by their area and SBH, and the
corresponding characteristic parameter is described by the Gaussian
distribution~\cite{Tung}. It was shown in a number of works
\cite{Iucolano,Iucolano2} that those theories can be used together. In
the case
where inhomogeneous patches do exist, relation~(\ref{13m}) remains valid,
but
the quantity $\Phi _{b}^{0}$ means the SBH magnitude in the uniform
region.

In the literature \cite{Tung,Sarpatwari,Schmitsdorf}, the dependence
between the $\Phi _{b}$- and $n$-values obtained from the CVC
analysis was shown to be linear in the case of contacts with local
inhomogeneities. Moreover, $\Phi _{b}=\Phi _{b}^{0}$ at $n=n_{\rm
if}$, where
\begin{equation}
n_{\rm if}=1+\frac{1}{4}\left[ \frac{q^{3}N_{D}}{8\pi
^{2}\varepsilon _{s}^{3}\varepsilon _{0}^{3}V_{bb}^{3}}\right]
^{1/4} \label{11m}
\end{equation}%
is the ideality factor that takes the influences of image forces
into account \cite{Sarpatwari}, $V_{bb}=(\Phi _{b}^{0}-V_{n}-V)$ is
the band bending in the semiconductor layer near the contact,
$\varepsilon _{s}$ is the semiconductor dielectric constant, and
$\varepsilon _{0}$ is the dielectric permittivity of vacuum. In the
dependence $\Phi _{b1}(n_{1})$, similarly to the previous cases, two
linear sections are observed with a cusp at $T\approx 225$~K (see
Fig.~\ref{Fig5}, curve {\it 2}). The extrapolation procedure gave
$\Phi _{bn,{\rm I}}^{0}=(0.646\pm 0.005)$~V for the interval
$T=(130\div 220)$~K and $\Phi _{bn,{\rm II}}^{0}=(0.64\pm 0.02)$~V
for the interval $T=(230\div 330) $~K.

\begin{table*}
\vskip3mm \noindent\caption{Determinated parameters for
Mo/\boldmath$n$-Si Schottky diodes}\vskip3mm\tabcolsep16.5pt
\noindent{\footnotesize
\begin{tabular}{|l| c| c| c| c|}
 \hline%
 {\parbox[c][8mm][c]{50mm}{\vspace*{7mm}Determination technique}}%
 & \multicolumn{2}{|c}{\rule{0pt}{5mm}Barrier height, V}
 & \multicolumn{2}{|c|}{Richardson constant, A/cm$^{2}$/K$^{2}$}\\[2mm]
\cline{2-5}%
\multicolumn{1}{|c|}{\rule{0pt}{5mm}}
 & \multicolumn{1}{|c}{130--220~K}
 & \multicolumn{1}{|c}{230--330~K}
 & \multicolumn{1}{|c}{130--220~K}
 & \multicolumn{1}{|c|}{230--330~K}\\[2mm]
\hline%
\rule{0pt}{5mm}Richardson dependence&0.141&0.599&$3.7\times10^{-10}$&32\\%
Dependence $\Phi_{b}$ versus $n$&0.646&0.64&&\\%
Dependence $\Phi_{b}$ versus $(kT)^{-1}$&0.872&0.656&&\\%
Modified Richardson dependence&0.874&0.655&125&110\\%
VFC&&0.689&&\\%
Source\cite{Schroder}&&&\multicolumn{2}{|c|}{112}\\[2mm]%
\hline
\end{tabular}
}\label{tbl}
\end{table*}

Taking Eqs.~(\ref{2m}) and (\ref{13m}) into account, the Richardson
dependence for the case of a barrier with different inhomogeneous
patches
can be written down in a modified form \cite{Tascioglu,Yildirim}
\begin{equation}
\ln\left(\! \frac{I_{\rm S}}{T^{2}}\!\right) -\left(\!
\frac{q^{2}\sigma_{\Phi}^{2}}{2k^{2}T^{2}}\!\right)
=\ln(SA^{\ast})-\frac{q\Phi_{b}^{0}}{kT}.   \label{16m}
\end{equation}
The corresponding plots for the obtained $\sigma_{\Phi,{\rm I}}$-
and $\sigma _{\Phi,{\rm II}}$-values are shown in Fig.~\ref{Fig6}.
The linear approximations of those curves in the corresponding
temperature intervals selected for the determination of
$\sigma_{\Phi,{\rm I}}$ and $\sigma_{\Phi,{\rm II}}$ brought about
the following parameters: $\Phi_{bRM,{\rm I}}^{0}=(0.874\pm0.004)$~V
and $A_{RM,{\rm I}}^{\ast}=(125\pm$\linebreak
$\pm20)$~A$/$cm$^{2}$/K$^{2}$ for $T=(130\div220)$~K, and
$\Phi_{bRM,{\rm II}}^{0}=(0.655\pm0.003)$~V and $A_{RM,{\rm
II}}^{\ast}=(110\pm$ $\pm10)$~A$/$cm$^{2}$/K$^{2}$ for
$T=(230\div330)$~K. Note that the magnitudes of $A_{RM,{\rm
I}}^{\ast}$ and $A_{RM,{\rm II}}^{\ast}$ practically coincide with
the corresponding literature data within the measurement errors. The
parameters obtained in different ways are listed in Table~1.

The temperature dependence of the ideality factor is known to depend
on the charge transfer mechanism. For example, if the thermal field
emission (TFE) or deep-level-assisted tunneling dominates, then
\cite{Rhoderick,Evstropov}
\begin{equation}
n={E_{00}}/{(kT)}\coth \left[ {E_{00}}/{(kT)}\right],
\label{14m_new}
\end{equation}%
where ${E_{00}}$ is a characteristic energy. Note that, in the TFE
case, $E_{00}=(\hbar /2)[N_{D}/(m^{\ast }\varepsilon _{s}\varepsilon
_{0})]^{1/2}$, where $m^{\ast }=1.08\times 9.11\times 10^{-31}$~kg
is the effective electron mass; therefore, for the examined
specimens and the temperature interval, there would have been
$n\approx 1$ in this case. However, if TE prevails, the temperature
dependence of $n$ for real contacts is often written down in the
form \cite{Rhoderick}
\begin{equation}
n=1+{T_{0}}/T,  \label{14m}
\end{equation}%
where $T_{0}$ is a certain constant. In the case of different patches,
it
was shown \cite{Sullivan,Tung,Iucolano} that
\begin{equation}
T_{0}={q\sigma _{\Phi }^{2}}/{(3kV_{bb})}.  \label{15m}
\end{equation}

\begin{figure}[b]%
\vspace*{-4mm}
\includegraphics[width=\column]{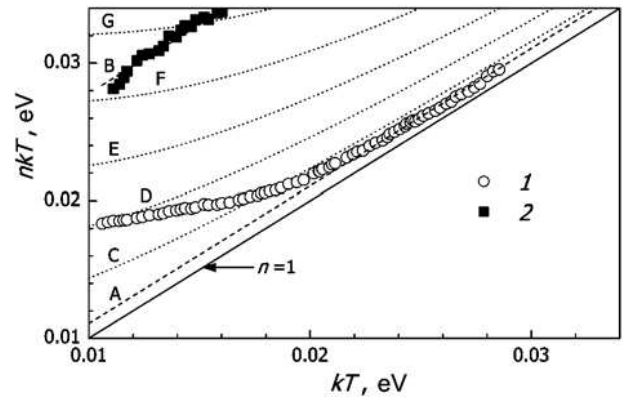}
\vskip-3mm\caption{Temperature dependences of the inverse CVC slope
$n_{1}$ (\textit{1}) and $n_{2}$ (\textit{2}). Dotted curves exhibit
the results of theoretical calculations according to formulas
(\ref{14m}) (curves A and B) and (\ref{14m_new}) (curves C to G).
$T_{0}=12$ (A) and 206$~\mathrm{K}$ (B). $E_{00}=12$ (C), 17 (D), 22
(E), 27 (F), and 32~mV (G). Solid straight line corresponds to the
ideal case $n=1$ }\label{Fig7}\vspace*{1.5mm}
\end{figure}

In Fig.~\ref{Fig7}, the calculated dependence for the inverse CVC
slope, $nkT$, and a number of curves calculated by
Eqs.~(\ref{14m_new}) and (\ref{14m}) are depicted. One can see that
the data obtained for $n_{1}$ at high temperatures are described
satisfactorily by expression (\ref{14m}) with
\mbox{$T_{0}=12~\mathrm{K}$}. On the other hand, the calculations by
Eq.~(\ref{15m}) and with the use of the obtained $\Phi _{bT,{\rm
II}}^{0}$- and $\sigma _{\Phi,{\rm II}}$-values show that, in the
temperature interval 230--330$~\mathrm{K}$, the model with local
inhomogeneities brings about rather a close value
$T_{0,\mathrm{theory}}\approx 11~\mathrm{K}$.

\begin{figure}
\includegraphics[width=7.8cm]{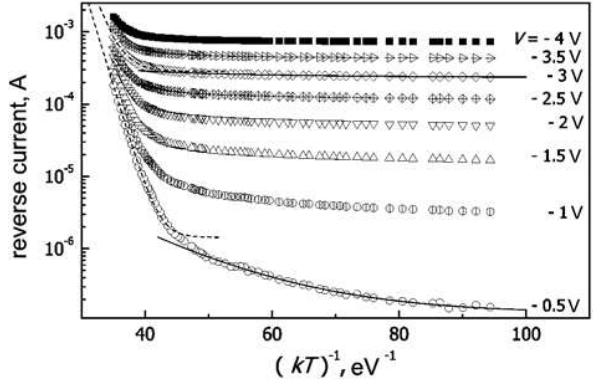}
\vskip-3mm\caption{Temperature dependences of the reverse current in
Mo/$n$-Si Schottky diodes at various bias voltages. Points
correspond to the experiment data, lines to their approximations by
formula (\ref{18m}) in the intervals $T=(130\div220)~\mathrm{K}$
(solid lines) and $T=(230\div330)~\mathrm{K}$ (dotted lines)
}\label{Fig8}
\end{figure}

\begin{figure}
\includegraphics[width=7.8cm]{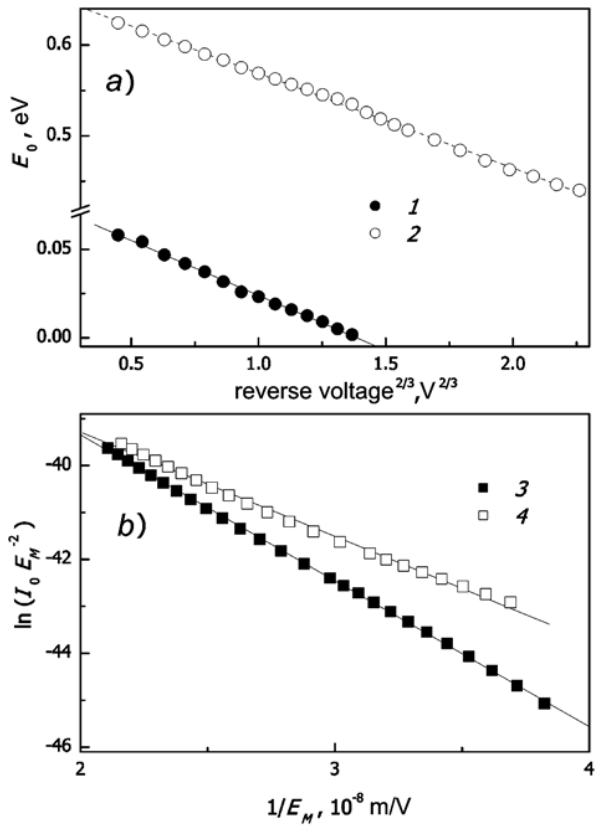}
\vskip-3mm\caption{Field dependences of the characteristic energy
(\textit{a}) and the temperature-independent component of the
reverse current in the Fowler--Nordheim coordinates (\textit{b}) at
$T=(130\div220)~\mathrm{K}$ (curves \textit{1} and \textit{3}) and
$T=(230\div330)~\mathrm{K}$ (curves \textit{2} and \textit{4}).
Points correspond to experimental data, straight lines to their
linear approximations  }\label{Fig9}
\end{figure}

Hence, the results presented above testify that the current $I_{1}$
can be described in the framework of the model considering TE
through a non-uniform barrier. An additional argument in favor of
this conclusion is a qualitative coincidence of the conventional and
transformed Richardson dependences (Fig.~\ref{Fig4}) and the
temperature dependence of $n_{2}$ in the interval
130--220~$\mathrm{K}$ (Fig.~\ref{Fig7}) with the corresponding
dependences predicted in the framework of this model (see
Fig.~11,{\it b} in work \cite{Tung} and Fig.~\ref{Fig3} in work
\cite{Sarpatwari}). By the way, note that, in the case where the SBH
is determined from VFCs, the influence of inhomogeneities is
insignificant \cite{Yildirim,Chen}. Therefore, the $\Phi _{b,{\rm
CV}}$-values can be compared with $\Phi _{b}^{0}$-ones obtained from
CVCs.

The only thing that needs a more detailed attention is the
difference between the $\Phi _{b}^{0}$- and $\sigma _{\Phi }$-values
in different temperature intervals, which is not predicted in the
framework of the theory for non-uniform contacts. At the same time,
we note that a similar situation has already been observed in
practice, e.g., in works \cite{Tascioglu,Yildirim,Mamor,Jiang}. Such
phenomena were explained as a domination of other, different from
TE, mechanisms of charge transfer at low temperatures, such as
TFE~\cite{Tascioglu,Mamor}, tunneling~\cite{Yildirim}, and
recombination processes~\cite{Mamor}. However, in our opinion, the
coincidence of the values obtained for $A_{R}^{\ast }$ with the
literature data testifies that it is just the TE theory that is
applicable to this case. The changes in the dependence slope in
Fig.~\ref{Fig5} can be associated with increase in the electron
emission rate by defects at the MS interface. Really, the level
depletion in some defects at $T\approx 225$~K should stimulate a
reduction of SBH and start a mechanism, owing to which some
inhomogeneous patches with elevated concentrations of similar
defects cease to be the regions of facilitated current passage due
to the effective capture of drifting electrons by traps. As a
result, $\sigma _{\Phi}$ has to diminish in the high-temperature
interval, which is really observed experimentally.

Now, let us come back to the current prevailing at low biases in the
low-temperature interval; it is $I_{2}$. In work~\cite{Tung}, it was
shown that such a current can appear at low temperatures in
non-uniform contacts as an additive to $I_{1}$ owing to the passage
of charge carriers through inhomogeneous regions. The ideality
factor for the corresponding CVC section should expectedly and
considerably exceed 1 at that, and a substantial influence of the
series resistance should also be observed. Just this phenomenon was
revealed in our researches (see Figs.~\ref{Fig1} and \ref{Fig3}). In
the case where the current through the patches is governed by the TE
mechanism, we have
\begin{equation}
I_{\rm S}=Sf_{p}A^{\ast }T^{2}\exp \left[ -{q\Phi
_{b,p}}/{(kT)}\right], \label{17m}
\end{equation}%
where $f_{p}$ is a multiplier that takes the area of
inhomogeneous patches into account, and $\Phi _{b,p}$ is the average value of SBH
in those regions. As is seen from Fig.~\ref{Fig3}, the SBH $\Phi
_{b2}$, being calculated on the basis of formula (\ref{2m}), is a
linear function of the temperature, $\Phi _{b2}=a_{\Phi }+b_{\Phi
}T$, where $a_{\Phi }=(0.056\pm 0.001)$~V and $b_{\Phi }=2.4\times
10^{-3}$~V/K. Comparing Eqs.~(\ref{2m}) and (\ref{17m}), we may
write down that $f_{p}=\exp (-qb_{\Phi }/k)\approx 10^{-12}$ and
$\Phi _{b,p}=a_{\Phi }=0.056$~V. As concerns the quantity $n_{2}$,
its temperature dependence is also described well by formula
(\ref{14m}) with $T_{0}=(206\pm 5)~\mathrm{K}$ (Fig.~\ref{Fig7}).

In Fig.~\ref{Fig8}, the dependences of the reverse current in
examined structures on the reciprocal temperature in the bias
interval $V=-(0.5$--$4.0)~\mathrm{V}$ are exhibited. It was found
that two temperature sub-intervals, 130--220 and
230--330$~\mathrm{K}$, are also expedient to be considered in this
case. In each of them, the temperature dependence of the reverse
current at a constant voltage is well approximated by the expression
\begin{equation}
I=CT^{2}\exp \left[ -{E_{0}}/{(kT)}\right] +I_{0},  \label{18m}
\end{equation}%
where the first term describes the TE current component, and the second one is
the
temperature-independent one, with the parameters $C$ and $E_{0}$ being
also
independent of the temperature.

The revealed dependence of the characteristic energy $E_{0}$ on the applied
voltage evidences the variation of SBH. It is known
\cite{Rhoderick,Tung,Andrews} that a reduction of the barrier height, when
the
reverse bias is applied, can occur owing to the influence of image forces
(in
this case, the SBH variation is $\Delta \Phi _{b}\sim V^{1/4}$) and
the electric
field ($\Delta \Phi _{b}\sim V^{1/2}$), as well as to the influence of
inhomogeneous regions. In the latter case, $\Delta \Phi _{b}\sim
V^{2/3}$,
and the proportionality coefficient depends on the local patch
parameters
\cite{Tung}. For the structures concerned, the quantity $E_{0}$ acquires
different values in every temperature interval. However, if the reverse
bias
increases, $E_{0}$ decreases as a linear function of $V^{2/3}$ in both
cases
(Fig.~\ref{Fig9},$a$). Hence, the analysis of the reverse CVC branches also
confirms that the current through the analyzed structures can be described
in
the framework of the model for a non-uniform contact with patches, the
influence of which on the charge transfer changes at a temperature of
about
225$~\mathrm{K}$.

The relative contribution of the temperature-independent component
to the reverse current was found to grow with the bias voltage. In
Fig.~\ref{Fig9},$b$, the field dependences of the current $I_{0}$
are depicted in the Fowler--Nordheim coordinates,
$\ln(I_{0}/E_{M}^{2})$ versus $(1/E_{M})$, where
$E_{M}=[2qN_{D}V_{bb}/(\varepsilon_{s}\varepsilon_{0})]^{1/2}$ is
the electric field strength at the metal--semiconductor interface
\cite{Rhoderick}. While calculating $E_{M}$, we used the
$\Phi_{bT}^{0}$-value obtained for the corresponding temperature
interval and the $V_{n}$-value averaged over it. The linear behavior
of dependences in Fig.~\ref{Fig9},$b$ and the independence of the
current component $I_{0}$ of the temperature testify to the tunnel
origin of this \mbox{component}.\looseness=1

\section{Conclusions}

In this  work, we have experimentally studied the forward and
reverse CVC branches for Mo/$n$-Si structures with a Schottky
barrier in the temperature interval 130--330~K. We have found that
the barrier height increases with the temperature, whereas the
ideality factor demonstrates the opposite tendency. It is shown that
the obtained results can be explained in the framework of the model
of thermionic emission through a contact containing local regions
with lowered values of barrier height. The barrier height in the
uniform contact region and the standard deviation of SBH are
determined to be 0.872 and 0.099~V, respectively, in the interval
130--220$~\mathrm{K}$, and 0.656 and 0.036~V, respectively, in the
interval 230--330$~\mathrm{K}$. For the transformed Richardson
dependence, we determined the Richardson constant, $115\pm
10$~A$/$cm$^{2}$/K$^{2}$. The average barrier height in the
inhomogeneous regions was determined to be 0.056~V. The reverse
current was demonstrated to be driven by both thermionic emission
through the non-uniform barrier and tunneling, with the relative
contribution of the latter mechanism growing if the bias voltage
increases.

\vskip3mm {\it The author is grateful to A.B.\,\,Nadtochii for his
assistance in carrying out VFC measurements.}


\rezume{О.Я. Оліх}{ОСОБЛИВОСТІ ПЕРЕНЕСЕННЯ ЗАРЯДУ\\ В СТРУКТУРАХ
Mo/$n$-Si З БАР'ЄРОМ ШОТКИ} {У роботі експериментально досліджено
прямі та зворотні вольт-амперні характеристики структур Mo/$n$-Si з
бар'єром Шотки в діапазоні температур 130--330~К. Виявлено, що при
підвищенні температури має місце збільшення висоти бар'єра Шотки та
зменшення фактора неідеальності. Проведено аналіз отриманих
результатів у рамках моделі неоднорідного контакту. Визначено
середнє значення та стандартне відхилення висоти бар'єра Шотки:
0,872~В та 0,099~В при $T=130$--220~К і 0,656~В та 0,036~В при $T =
230$--330~К відповідно. Показано, що при зворотному зміщенні
основними процесами перенесення заряду є\linebreak термоелектронна
емісія через неоднорідний бар'єр та тунелювання.}

\end{document}